# π−π Stacking between Polyaromatic Hydrocarbon Sheets beyond Dispersion Interactions

Nadeesha J. Silva[1], Francisco B. C. Machado[2], Hans Lischka*[1,3,4], Adelia J. A. Aquino*[1,4,5]

[1]Department of Chemistry and Biochemistry, Texas Tech University, Lubbock Texas 79409, United States

[2]Departamento de Química, Instituto Tecnológico de Aeronáutica, São José dos Campos, 12.228-900 São Paulo, Brazil

[3]Institute for Theoretical Chemistry, University of Vienna, A-1090 Vienna, Austria

[4]School of Pharmaceutical Sciences and Technology, Tianjin University, Tianjin, 300072 P.R. China

[5]Institute for Soil Research, University of Natural Resources and Life Sciences Vienna, A-1190 Vienna, Austria

E-mail:
Adelia J. A. Aquino: adelia.aquino@univie.ac.at
Hans Lischka: hans.lischka@univie.ac.at

Abstract

High level *ab initio* calculations ranging from coupled cluster methods including explicitly correlated approaches to standard second order Møller-Plesset theory using spin scaling (SOS-MP2) have been performed on sandwich and slipped parallel dimer structures of a series of quasi one-dimensional acenes and on two-dimensional sheets containing the series pyrene to coronene encircled with two layers of benzene rings. Sandwich (graphitic AA type) and slipped parallel (AB type) structures were considered and, within given symmetry restrictions, full geometry optimizations were performed. Basis set superposition effects have been considered. The computed geometries show a significant biconcave deviation of the two-dimensional sheets from planarity with the central intersheet C···C distances considerably smaller that van der Waals distances. The computed intersheet binding energy per carbon atom extrapolated for $N \to \infty$ of -74.3 meV (1.713 kcal/mol)/atom agrees quite well with the experimental defoliation energy of -52 meV (1.199 kcal/mol)/atom (-67 meV (1.545 kcal/mol)/carbon atom without corrections for H binding contributions) for polyaromatic hydrocarbons (PAHs) from graphite. A limited investigation of density functional theory (DFT) calculations using empirical dispersion contributions has been performed also showing a significant underbinding character of the D3 method. For most of the DFT variants investigated the graphene sheet models retain a quasi-planar structure in strong contrast to the afore-mentioned SOS-MP2 results.

# 1 Introduction

Understanding the nature of π-π stacking interactions of conjugated π systems is of crucial importance in many areas of chemistry, molecular biology and material science. They are the primary non-covalent interactions that influence base stacking in DNA and RNA,[1, 2] determining chemical and biological recognition processes,[3] and are of crucial importance in the field of materials science to understand noncovalent interactions of carbon nanostructures.[4] From specific interesting applications in the latter field we want to mention the exfoliation of graphene from graphite[5-7] where one promising procedure to prevent re-aggregation of the dissolved graphene layers consists in the adsorption of smaller functionalized polycyclic aromatic hydrocarbons (PAH) on graphene layers[8] to stabilize individual sheets in solution. In view of the importance and the mentioned wide-spread occurrence of stacked π-π interactions a thorough understanding and the ability for reliable predictions of the respective interaction energies and corresponding structures of the interaction complexes by means of theoretical simulations is highly desirable. However, accurate quantum chemical calculations of van der Waals (vdW) interactions still constitute significant computational challenges.[9, 10] Most of the commonly used functionals in DFT lack dispersion interactions[11] which inhibits a straightforward application of these methods to van der Waals complexes. Density functional theory (DFT) calculations augmented with empirical dispersion terms has been used to overcome this problem.[12-18] Within this frame several procedures have been suggested such as the DFT-D [12, 13, 15, 19] and DFT+vdW[16] methods based on a sum of $C_6R^{-6}$ terms over atom pairs or the vdW-DFT method[20] to resolve this problem (for a review see e.g. Ref. [21]). From the available dispersion correction methods, the DFT-D method[12, 13, 15, 19] provides a pragmatic, but computationally efficient and convenient approach which has been successfully applied among others to study π-stacked adsorption processes on graphene sheets.[22-25]

On the other hand, a larger selection of *ab initio* methods is available for accurate calculation of dispersion energies. Because of its relatively small size, the benzene dimer has been investigated intensively at several levels of sophistication focusing on coupled cluster methods with singles and doubles and non-iterative triples (CCSD(T)) which are believed to give the most reliable results at present.[26-32] Basis set extrapolation and explicitly correlated (F12) methods[33] provide the basis for the most reliable calculations.[29, 30, 34-37] These studies show the existence of two almost isoenergetic structures, a T-shaped and a parallel-displaced (PD) one (Scheme 1) with interaction energies around -2.5 kcal/mol with the T-shaped structure slightly more stable. Both structures are more stable than the S (sandwich) structure.[28-30] It should be noted, however, at this point that the PD structures show higher stability over the T shaped types in the larger stacked acene dimers.[38, 39]

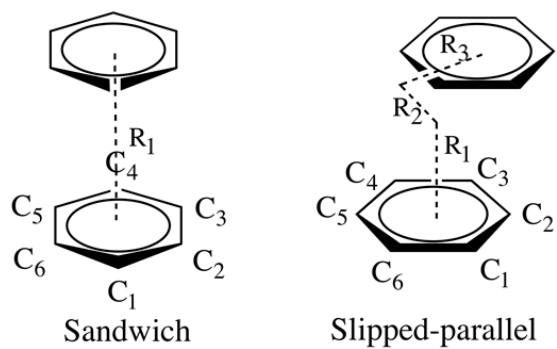

Scheme 1. Sandwich and slipped-parallel structures of benzene. $R_1$ is perpendicular to the benzene rings, $R_2$ is along $C_1$-$C_4$ and $R_3$ is perpendicular to $R_2$.

In comparison to the detailed calculations on the benzene dimer described above, only few coupled cluster calculations exist on larger PAH dimers because of the drastic increase in computational cost. Using reduced basis set sizes as compared to the benzene dimer, CCSD(T) calculations have been performed on the naphthalene[40-42] and coronene[40, 43] dimers. Accurate experimental results from Molecular Beam-Laser Spectroscopy measurements and high-level CCSD(T) calculations for the stacking interaction of the anisole dimer should be mentioned in the present context as well.[44]

The high computational cost of the CCSD(T) method precludes its use for significantly larger systems than the benzene or naphthalene dimers. Thus, cheaper *ab initio* alternatives are of high interest, from which the second order Møller-Plesset perturbation theory (MP2)[45] is certainly one of the most attractive alternatives. However, it has been found that MP2 significantly overestimates the interaction energy for the benzene dimer in comparison to CCSD(T) results.[27, 28, 46] To correct for this deficiency, a spin component scaling (SCS) method has been introduced by Grimme[47] for MP2. Because the SCS scaling factor was much smaller for the same spin part as compared to the scaling factor in the opposite spin part, Jung et al.[48] noticed that neglecting completely the same spin part would have the same effect leading to the scaled opposite-spin (SOS) method. It has been shown by Hill et al.[49] by means of local MP2 (LMP2) SCS calculations on the benzene dimer, that the SCS approach was well suited for the accurate description of van der Waals interactions. This finding has been confirmed by systematic investigations performed by Antony and Grimme[50] on a larger test set. There it is also noted that basis set superposition error and basis set incompleteness almost cancel at a triple-zeta quality level basis set. The good performance of the SOS-MP2 was also shown for the stacked interactions of tetracyanoethylene with benzene, naphthalene and anthracene, respectively,[51] and on the much larger fullerene-porphyrin complex.[52]

Because of the just-described methodological problems and the large computational effort involved in high-level *ab initio* methods it is still difficult to find reliable but computationally efficient methods for calculating stacking interactions between larger PAHs. From the discussion above it appears that two methods are of practical interest in this context, these are the SOS-MP2 method to be used for benchmarking purposes and DFT-D for even more extended molecular systems. In the present work we chose as starting point our own CCSD(T), CCSD(T)-(F12) and SOS-MP2-(F12) calculations where the explicitly correlated methods are of particular interest since they improve the basis set convergence considerably[33] and are especially useful for highly accurate calculations intermolecular interactions.[34, 51, 53] In addition to the choice of the adequate computational method the basis set superposition error (BSSE) poses a major problem which is usually addressed by means of the Boys-Bernardi counterpoise approach.[54] Numerous investigations have addressed this problem[55-59] which led to various procedures in order to cope with the frequently observed underbinding of the BSSE corrected interaction energies as opposed to the commonly found overbinding of the uncorrected values. Basis set extrapolation to the complete basis set limit[60] based on explicitly correlated methods[61, 62] facilitates the analysis of the BSSE considerably. The just-mentioned over-/underbinding of interaction energies has led to the suggestion of using their average[59] whose success has been explained by Brauer et al.[61] on the basis of error compensation between BSSE and intrinsic basis set completeness.

Unfortunately, there is no unique black-box scheme available and actual procedures have to be tested and selected for special classes of applications. In the present work we use high-level methods for stacked quasi one-dimensional acene dimers in the spirit of the BSSE investigations described in the previous paragraph as a preparation for subsequent calculations on dimers containing two-dimensional sheets. In pursuing this goal, one should not forget that the size of the graphene sheets to be investigated is quite large (up to 300 carbon atoms as in the case of the coronene circum-3 dimer (for definition see below)) at least for the *ab initio* methods to be used. Thus, our goal is to develop a reasonable balance between accuracy and feasibility of such calculations and to show a feasible path for performing quantum chemical calculation of stacking energies of dimers of graphene nanosheets.

## 2 Computational Details

Geometry optimizations have been performed using the SOS-MP2[48] method and the DFT based on the hybrid Becke, 3-parameter, Lee-Yang-Parr (B3LYP),[63] the gradient-corrected correlation functional of Perdew, Burke and Ernzerhof (PBE)[64] including the D3 dispersion correction of Grimme et al.[15] Additionally, the B97 functional in combination with the D2 dispersion contribution[13] has been used for comparison. Unless stated differently, both sandwich and slipped parallel geometries of all investigated structures were fully optimized. Interaction energies were calculated with respect to the sum of the energies of the isolated monomers. The split valence, SV(P)[65] basis and the triple zeta valence polarization def2-TZVP[66] basis set containing two d and one f polarization set on carbon were chosen for the present investigations. The resolution of identity (RI) or density-fitting framework[66-68] has been used in connection with the SOS-MP2 and DFT/PBE methods.

Slipped-parallel structures were generated as shown in Scheme 2 for the naphtalene and pyrene dimers, respectively.

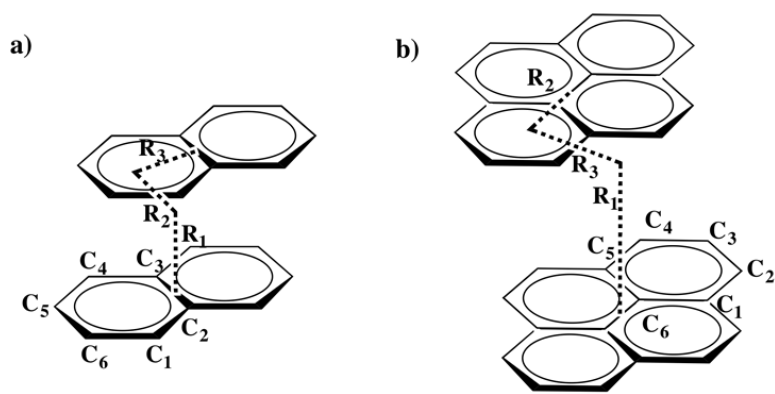

Scheme 2. Slipped-parallel structure of a) naphthalene and b) pyrene dimer. $R_1$ is perpendicular to the PAH sheets, $R_2$ is along $C_2$-$C_3$ (naphthalene) and goes from the center through $C_3$ (pyrene), respectively, and $R_3$ is perpendicular to $R_2$.

In addition to the DFT and SOS-MP2 methods used for geometry optimization, single point calculations were performed with the CCSD(T) method for four of the dimers (benzene, naphthalene, anthracene and pyrene) and potential energy curves were obtained as a function of the distance between the two monomers of the sandwich type configurations. The inter-monomer distance (R) was symmetrically varied around the energy minimum and the CCSD(T) data were fitted to a 4$^{th}$ degree polynomial function. Furthermore, to obtain a better estimate of basis set effects, the explicitly correlated SOS-MP2-F12[69] and CCSD(T)-(F12*)[70] approaches (for a review see also Ref. [71]) were used in single point calculations using the cc-pVDZ-F12 basis. The Turbomole suite of programs[72] was used in all calculations.

## 3. Results and Discussion

3.1 Acene Dimers

Interaction energies and interring distances were calculated for the stacked sandwich structures of linear acenes (Scheme 2) by means of the CCSD(T), SOS-MP2 and DFT/B3LYP-D3 methods. Results obtained for the benzene dimer are collected in Table 1 and Table 1S of the Electronic supplementary information (ESI). Inspection of Table 1 shows the expected fact that the MP2 method largely overestimates the binding energy. The SOS-MP2 variant, however, performs significantly better in comparison with higher level results also displayed in this table. While the interaction energy obtained with the smaller SV(P) basis underestimates the interaction energy somewhat (Table 1S), the significantly larger def2-TZVP basis yields interaction energies quite close to the benchmark CCSD(T) and CCSD-F12*(T) results. The SOS-MP2-F12 approach, however, decreases the interaction again. Except for the MP2 results, the optimized interring bond distance R shows a relatively small variation with the method used. Comparison with CCSD results which exclude triples contributions shows the importance of the latter. The CCSD method gives an interaction energy of only -0.999 kcal/mol as compared to -1.841 kcal/mol for CCSD(T). Curve fitting in the intermolecular distance $R$ has been performed for selected methods. Its effect on $R$ and the interaction energy is quite modest. The most accurate value of 3.900 Å is obtained from a curve fitting of CCSD*(T)/cc-pVDZ-F12 data and compares well with the CCSD(T)/CBS value of 3.89 Å reported in Ref. [31] The DFT/B3LYP-D3 interaction energy is located well within the range of the *ab initio* results. With 3.81 Å the interring distance is notably on the lower end of values shown in Table 1.

**Table 1**. Interaction energies $\Delta E$, inter monomer distances $R$ of the optimized stacked dimers of benzene using several methods and basis sets.[a]

| Method [a] | Basis | $\Delta E$ (kcal/mol) | $R$ (Å) |
|---|---|---|---|
| MP2 | def2-TZVP | -3.523 | 3.799 |
| SOS-MP2 | def2-TZVP | -1.563 | 3.931 |
| SOS-MP2-F12[b] | cc-pVDZ-F12 | -1.340 | 3.931 |
| SOS-MP2-F12[c] | cc-pVDZ-F12 | -1.343 | 3.956 |
| CCSD | def2-TZVP | -0.999 | 3.931 |
| CCSD(T)[b] | def2-TZVP | -1.841 | 3.931 |
| CCSD(T)[c] | def2-TZVP | -1.852 | 3.879 |
| CCSD-F12*(T)[b] | cc-pVDZ-F12 | -1.828 | 3.931 |
| CCSD-F12*(T)[c] | cc-pVDZ-F12 | -1.831 | 3.900 |
| DFT/B3LYP-D3 | def2-TZVP | -1.690 | 3.814 |
| Previous work | | | |
| CCSD(T)/CBS[26] | - | -1.70 | 3.9 |
| CCSD(T)/CBS[31] | - | -1.66 | 3.89 |

[a] Geometries were optimized with the same method as used for the energy calculations unless specified differently; [b] SOS-MP2/def2-TZVP geometry; [c] from the potential energy curve fitting in the distance $R$. Remaining geometry fixed from SOS-MP2/def2-TZVP geometry.

The most extended data for the stacked dimers of naphthalene and anthracene using the def2-TZVP basis are listed in Table 2. Results for the SV(P) basis are given in Table 1S. The CCSD(T)/def2-TZVP interaction energy is 0.3 kcal/mol stronger than the corresponding SOS-MP2 result; SOS-MP2-F12 reduces the interaction by 0.7 kcal/mol as compared to SOS-MP2. Adding this basis set effect to the CCSD(T)/def2-TZVP interaction energy of -4.773 kcal/mol leads to our best estimate of -4.1 kcal/mol. The DFT-D3/def2-TZVP result of -4.216 (or -3.909 kcal/mol with BSSE) fits well to our *ab initio* results.

For the anthracene dimer CCSD(T) calculations were restricted to the SV(P) basis since using the def2-TZVP basis exceeded our computational capabilities. Comparison of SV(P) results (Table 2 and Table 1S) shows an increase in the interaction energy from SOS-MP2 to CCSD(T) by 0.375 kcal/mol in absolute value. Adding this increment to the result computed with the SOS-MP2-F12 method (Table 2) leads to our best estimate for the interaction energy of -6.8 kcal/mol. The DFT/B3LYP-D3/def2-TZVP result is very close, but is reduced to -6.449 kcal/mol by the BSSE correction to be discussed further below.

For the tetracene and pentacene dimers SOS-MP2 and DFT/B3LYP-D3 results are collected in Table 1S. The pattern observed for the SOS-MP2 calculations is the following: increasing the basis set from SV(P) to def2-TZVP enhances the interaction energy in the range of 1-2 kcal/mol. Application of the SOS-MP2-F12/cc-pVDZ-F12 approach reduces this interaction by ~2 kcal/mol. CCSD(T) calculations were not feasible any more with the computational resources available. However, the experience obtained from the smaller acene dimer cases indicates that they would again enhance the interaction again in a similar range of 1-2 kcal/mol. This analysis shows a significant occurrence of error compensation which could finally give support to results obtained with the SOS-MP2 method and smaller basis sets. This is especially of interest for significantly larger stacked systems to be discussed in the following sections. The interaction energies computed by the DFT/B3LYP-D3/def2-TZVP method are located at the lower end of the energy scale established by the SOS-MP2 results (Table 2). Consequently, the interchain distances are too long by 0.1 Å (def2-TZVP result).

**Table 2**. Interaction energies ΔE and inter monomer distances $R$ of the optimized stacked dimers of naphthalene and anthracene using different basis sets.

| Method [a] | Basis set | $\Delta E$ (kcal/mol) | $R$ (Å) |
|---|---|---|---|
| Naphthalene dimer | | | |
| SOS-MP2 | def2-TZVP | -4.437 | 3.800 |
| SOS-MP2-F12[b] | cc-pVDZ-F12 | -3.730 | 3.800 |
| SOS-MP2-F12[c] | cc-pVDZ-F12 | -3.739 | 3.831 |
| CCSD[b] | def2-TZVP | -2.884 | 3.800 |
| CCSD(T)[b] | def2-TZVP | -4.767 | 3.800 |
| CCSD(T)[c] | def2-TZVP | -4.773 | 3.773 |
| Best estim.[d] | | -4.1 | - |
| DFT/B3LYP-D3 | def2-TZVP | -4.216 | 3.808 |
| | | -3.909[e] | |
| Anthracene dimer | | | |
| SOS-MP2 | def2-TZVP | -7.726 | 3.757 |
| SOS-MP2-F12[b] | cc-pVDZ-F12 | -6.461 | 3.757 |
| CCSD(T)[b] | SV(P) | -7.105 | 3.757 |
| Best estim.[d] | | -6.8 | |
| DFT/B3LYP-D3 | def2-TZVP | -6.832 | 3.805 |
| | | -6.449[e] | |

[a] Geometries were optimized with the same method as used for the energy calculations unless specified differently; [b] SOS-MP2/def2-TZVP geometry; [c] From the potential energy curve fitting in the distance R. Remaining geometry fixed from SOS-MP2/def2-TZVP geometry; [d] see text; [e] BSSE corrected

As already discussed in the Introduction, previous experience concerning BSSE corrections shows that the uncorrected interaction energy frequently overbinds whereas the BSSE correction leads to an underbinding interaction mechanism. Based on these findings, it has been recently argued by Burns et al.[59] to pragmatically use the average of both values. Since it is computationally not feasible to obtain truly converged counter-poise corrected interaction energies for the size of systems investigated here, we want to make use of the observations of Brauer et al.[61] that the cc-pVDZ-F12 basis in connection with MP2-F12 gives good uncorrected results due to an error compensation of overbinding BSSE and underbinding intrinsic basis set insufficiencies. Following this line of reasoning, in Table 3 uncorrected, BSSE corrected and average SOS-MP2/def2-TZVP values are compared to explicitly correlated SOS-MP2-F12 results. In relation to SOS-MP2(F12), the uncorrected and corrected SOS-MP2/def2-TZVP data show the expected over- and underbinding and a quite good agreement with the averages. Therefore, we take the average interaction energy as our best result in the subsequent calculation of the significantly larger 2-dimensional graphene sheet models where SOS-MP2-F12 calculations are not feasible anymore.

In Table 2S the BSSE analysis is presented for the SV(P) basis. The results show a strong underbinding of the BSSE corrected values and a relatively small overbinding of the uncorrected results which is, however, increasing with the size of the systems. The average is always underbinding. Table 3S shows the BSSE analysis for the DFT/B3LYP SV(P) and def2-TZVP approaches. In comparison with SOS-MP2 of Table 3 the DFT calculations show a significantly smaller BSSE correction. Because of the a-posteriori added dispersion term based on the D3 approach this term is not affected by basis set effects and, thus, it is not source for possible underbinding because of intrinsic basis insufficiency. Therefore, we consider the BSSE corrected values as our best DFT results.

**Table 3**. BSSE corrected, uncorrected and average interaction energies $\Delta E$ for the acene dimer series using the SOS-MP2/def2-TZVP approach in comparison with SOS-MP2(F12)/cc-pVDZ-F12 results, respectively. All values are in kcal/mol.

| Dimer | $\Delta E$uncorr.[a] | $\Delta E_{BSSE}$(corr.) | $\Delta E$(aver.) | $\Delta E$ |
|---|---|---|---|---|
| | SOS-MP2/def2-TZVP | | | SOS-MP2(F12)[b] |
| Benzene | -1.563 | -0.798 | -1.180 | -1.340 |
| Naphthalene | -4.437 | -2.706 | -3.572 | -3.730 |
| Anthracene | -7.726 | -4.990 | -6.358 | -6.461 |
| Tetracene | -11.256 | -7.427 | -9.341 | -9.410 |
| Pentacene | -14.908 | -9.990 | -12.449 | -12.608 |

[a] See Table 1 and Table 2; [b] cc-pDZV-F12 basis

The evolution of the interaction energies $\Delta E/N$ where $N$ is the number of C atoms in the acene monomer is displayed in Figure 1 for the SOS-MP2 and the DFT/B3LYP-D3 approaches. The $\Delta E/N$ values increase in absolute value with increasing chain length in both cases significantly. For the different SOS-MP2 variants (Figure 1a), def2-TZVP average, SV(P) uncorrected and the F12 curves agree closely. The averaged SV(P) interaction energies show a strong underbinding as already discussed above. The major difference of the DFT/B3LYP-D3 curves is a more pronounced flattening with increasing $N$ as compared to the SOS-MP2 case.

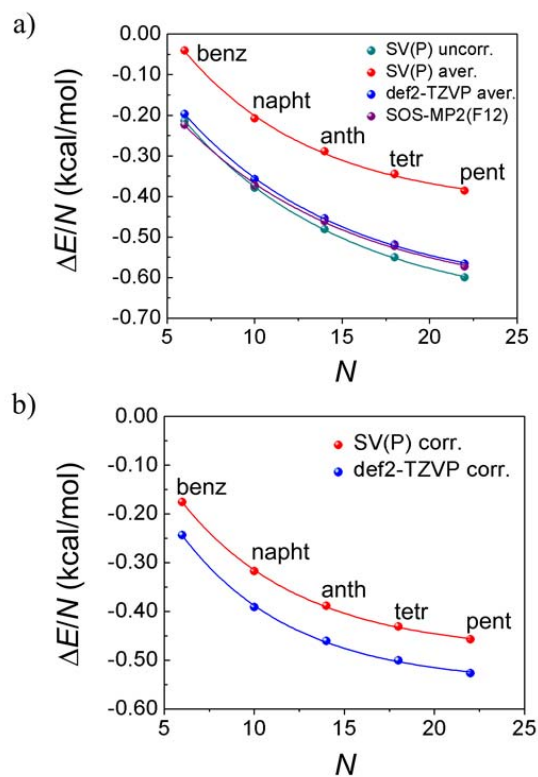

**Figure 1**. Interaction energies $\Delta E/N$ ($N$ is the numbers of C atoms in the monomer) for the sandwich structures of benzene to pentacene in dependence of $N$ for a) different SOS-MP2 versions and b) the DFT/B3LYP-D3 methods including BSSE corrections.

It has been shown that restricted Hartree-Fock and DFT calculations become triplet instable for *n*-acenes starting with $n = 7$[73] and that unrestricted symmetry-broken calculations result in lower energies. Following this approach, low-spin unrestricted SOS-MP2 (SOS-UMP2) and unrestricted DFT (UDFT) calculations have been performed for the monomer and stacked dimer of heptacene and decacene. The results are collected in Table 4S, Figure 1S and Figure 2S. The two figures show that both the restricted and unrestricteded approaches lead to an inconsistency in the $\Delta E/N$ curves and thus do not give a meaningful continuation of the interaction energies computed at spin-restricted level of the series up to the pentacene dimer. Thus, we did not investigate the interaction between larger acenes further.

Slipped parallel structures have been investigated for the series benzene to pentacene as well. They are displayed in Figure 2 for optimizations performed at the SOS-MP2/def2-TZVP level. The optimized structures show not only strong displacements from the sandwich structure but are also shifted from the AB-type graphitic structure[74] of the two-dimensional sheets discussed below. Interaction energies and displacement vectors are given in Table 5S. The evolution of the $\Delta E/N$ values of the slipped parallel acene dimer structures are shown in Figure 2S for the series benzene to pentacene. The pattern of the stabilization curves is similar to the one discussed for the sandwich structures. The stabilization of the interaction energy $\Delta E/N$ due to the slipping process amounts to ~0.3 kcal/mol for the pentacene dimer at SOS-MP2/def2-TZVP level (Table 1S and Table 5S). A smaller value of 0.15 kcal/mol is obtained for the benzene dimer.

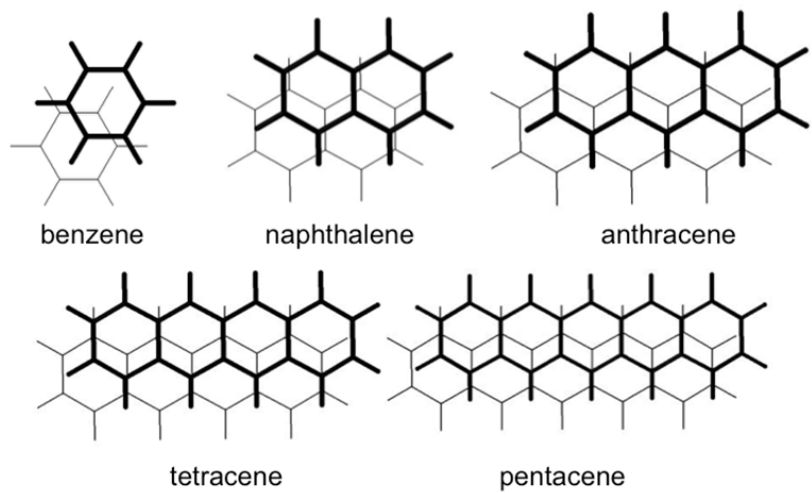

**Figure 2**: Top view of the slipped-parallel structures of the dimers of benzene to pentacene in $C_i$ symmetry using the SOS-MP2/def2-TZVP method.

3.2 Two-dimensional Sheets: Sandwich Structures

The sandwich structures of pyrene, perylene, coronene, hexabenzocoronene (HBC), coronene circum-1 and coronene circum-2 dimers (for monomers see Figure 3) will be discussed in this section. The coronene circum-1 and coronene circum-2 structures have been obtained by surrounding coronene with one and two sets of benzene rings, respectively.

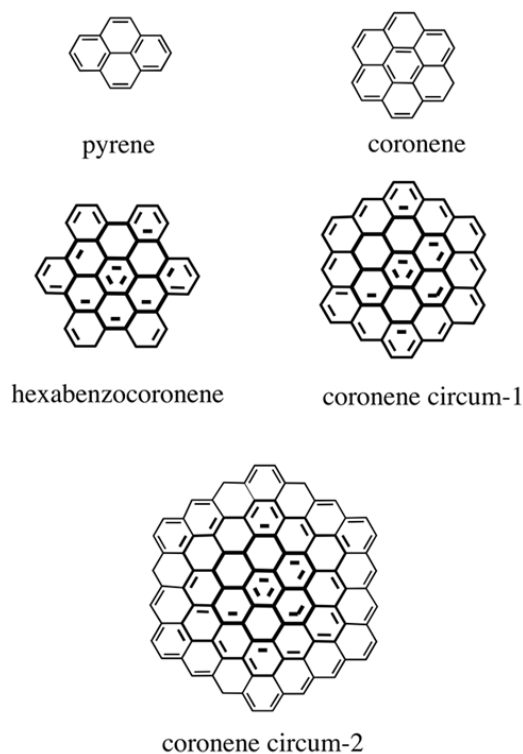

Figure 3. Structures used for stacked sandwich dimers of pyrene (16), coronene (24), hexabenzocoronene (42) and circular coronene extensions coronene circum-1 (54) and coronene circum-2 (96). Numbers in parentheses refer to the number of carbon atoms in the monomer.

Geometry optimizations have been performed in $D_{2h}$ symmetry. These structures are of interest because of their relation to AA stacking in graphite. It should be noted, however, that they correspond to second order saddle points as has been shown by Janowski et al.[43] in case of the sandwich structure of the coronene dimer. Interaction energies for the pyrene sandwich complex obtained with the SOS-MP2, SOS-MP2-F12, CCSD(T) and DFT/PBE-D3 approaches are collected in Table 4 and Table 6S. Increasing the basis set from SV(P) to def2-TZVP enhances the interaction energy by 1.3 kcal/mol at SOS-MP2 level whereas the SOS-MP2-F12 method leads to a reduction by 0.5 kcal/mol as compared to the SV(P) result. The average value of BSSE corrected and uncorrected SOS-MP2/def2-TZVP interaction energies (-8.663 kcal/mol, Table 7S) agrees very well with the SOS-MP2-F12 result of -8.556 kcal/mol (Table 4). Following the procedures used for the stacked naphthalene and anthracene dimers a best estimate of -8.8 kcal/mol is obtained by adding the increase in the interaction energy from SOS-MP2/SV(P) to CCSD(T)/SV(P) of 0.227 kcal/mol to the SOS-MP2-F12 result (Table 4). The BSSE corrected DFT/PBE-D3/def-TZVP interaction energy is ~0.9 kcal/mol lower (in absolute value) than the best estimate. The inter-sheet distance and the interaction energy show relatively little variation at CCSD(T) level when the intersheet distance is optimized.

**Table 4**. Interaction energies ΔE and inter monomer distances R of the pyrene sandwich complex using several methods and basis sets.

| Method[a] | Basis | ΔE (kcal/mol) | R (Å) |
|---|---|---|---|
| Pyrene dimer | | | |
| SOS-MP2 | SV(P) | -9.082 | 3.657 |
| SOS-MP2 | def2-TZVP | -10.371 | 3.704 |
| SOS-MP2-F12[b] | cc-pVDZ-F12 | -8.556 | 3.704 |
| CCSD(T)[b] | SV(P) | -9.309 | 3.704 |
| CCSD(T)[c] | SV(P) | -9.447 | 3.619 |
| Best estim. | | -8.921 | |
| DFT/PBE-D3 | SV(P) | -9.822 | 3.735 |
| DFT/PBE-D3 | def2-TZVP | -8.521 | 3.830 |
| | | -8.046[d] | |

[a] Geometries were optimized with the same method as used for the energy calculations unless specified differently; [b] SOS-MP2/def2-TZVP geometry; [c] from the potential energy curve fitting in the distance R. Remaining geometry fixed from SOS-MP2/def2-TZVP geometry; [d] BSSE corrected

The interaction energies $\Delta E/N$ for the sandwich structures of the systems pyrene to coronene circum-2 are displayed in Figure 4 for the SOS-MP2 method using BSSE uncorrected and average values. For the sake of completeness, the DFT/PBE-D3 results are shown in Figure 3S. Numerical values are collected in Tables 6S and 7S. As in the case of the stacked acene dimers (Figure 1), the SOS-MP2 interaction energy per C atom $\Delta E/N$ increases significantly in absolute value with the system size. Basis set effects follow the same pattern as discussed for the acene dimer series. In contrast to the situation found for the stacked acenes, the uncorrected Δ*E/N* values computed with the SOS-MP2/SV(P) method show a significant overbinding; the respective average interaction energies, however, agree quite well with the SOS-MP2/def2-TZVP results. Unfortunately, SOS-MP2 geometry optimization of the coronene circum-2 complex was not feasible at def2-TZVP level and single-point calculations using the SV(P) basis geometry did not result in consistent interaction energies. The DFT/PBE-D3 method shows a significant underbinding (Figure 3S). A more detailed discussion of the DFT/PBE-D3 results will be given for the slipped parallel structures below.

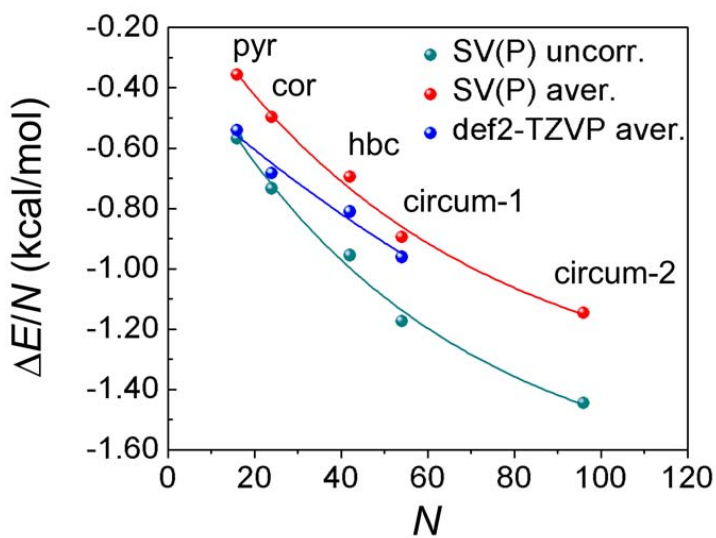

Figure 4. SOS-MP2 interaction energies Δ*E/N* (*N* is the numbers of C atoms in the monomer) for the sandwich dimers of pyrene to coronene circum-2 in dependence of *N* using the SV(P) and def2-TZVP basis sets.

The present results for the coronene dimer can be compared to the calculations of Janowski et al.[43] based on SCS-MP2 and quadratic configuration interaction with singles, doubles and perturbative triples (QCISD(T)) using modified augmented correlation consistent polarized double zeta (aDZ) and triple zeta (aTZ) basis sets and SAPT(DFT) calculations[75] with an aug-cc-pVDZ basis set. The SCS-MP2 and SAPT(DFT)/ interaction energies of -17.49 kcal/mol and -17.45 kcal/mol, respectively, compare well with our calculated range of -17.6 to - 19.7 kcal/mol at SOS-MP2 level (Table 7S). The QCISD(T)/aDZ interaction energy corrected for aTZ basis set effects is found to be -14.66 kcal/mol.

The interaction energies $\Delta E/N$ given in Figure 4 have been fitted by an exponential of the form

$$\Delta E/N = ae^{-N/b} + \Delta E_\infty \qquad \text{Eq. (1)}$$

in order to obtain the interaction energy $\Delta E_\infty$ extrapolated for $N \to \infty$. The quality of the fitting is measured by the coefficient of determination ($R^2$). At SOS-MP2 level only the SV(P) interaction energies could be extrapolated (Table 5) because of the lack of sufficiently large sheets. Good fits with $R^2$ values close to 1 were achieved in all cases.

As to be expected from the comparison of the shapes of the graphs presented in Figure 4 and Figure 3S, the DFT/PBE interaction energies are significantly smaller in absolute value by about a factor of two in comparison to the SOS-MP2 results.

**Table 5.** Extrapolated interaction energies $\Delta E_\infty$ for the two-dimensional sandwich structures and $R^2$ values.

| Method | $E_\infty$(kcal/mol)[a] | $R^2$ |
|---|---|---|
| SOS-MP2/SV(P) aver. | -1.444 (-62.6) | 0.986 |
| DFT/PBE BSSE corr. | | |
| SV(P) | -0.718 (-31.1) | 0.989 |
| def2-TZVP | -0.744 (-32.3) | 0.992 |

[a] Values in parentheses are given in meV.

In the following, the inter-sheet distances are analyzed in more detail. The sandwich dimers have a biconcave structure as shown for the coronene circum 2 structure in Figure 5. The dimer is considerably bent toward the center with an interring distance of 3.33 Å in this area which is significantly below the van der Waals distance of ~3.50 Å.[76] The edges show distances which are increased by ~0.3 Å.

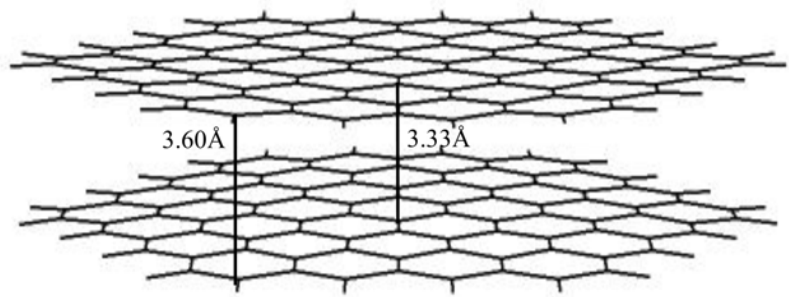

**Figure 5**. The biconcave structure of the coronene circum-2 dimer based on SOS-MP2/SV(P) optimization.

In Figure 6 the histograms including all perpendicular interring CC distances are presented as obtained from SOS-MP2/SV(P) calculations. SOS-MP2 results for the def2-TZVP basis set are given in Figure 4S. A significant shift of the distribution of interring CC distances toward smaller values is observed on increasing the size of the two-dimensional sheets. The SOS-MP2/SV(P) distribution starts with values around 3.65 Å for the pyrene dimer. The interring distances are successively reduced with increasing the size of the PAHs. The smaller distances always belong to the center of the ring systems and the larger ones to the edges. Beginning with HBC, the distribution of distances starts to move below the van der Waals distance of ~3.50 Å.[76] The minimal distances for coronene circum-2 are in the range of 3.35 – 3.40 Å. The evolution of the distances for the SOS-MP2/def2-TZVP approach (Figure 4S) looks quite similar. Most importantly, the lower portion of the distribution of distances moves well below the van der Waals distance region in this case also. These distances can be compared, for example, with the interring distances in the pancake bonds of the phenalenyl radical dimer of 3.1 Å[77] which also show a significant reduction in comparison to the van der Waals distance. It should be noted, however, that the shape of this dimer is typically biconvex in this case[77, 78] since the interactions due to the unpaired electron density is concentrated at the edges of the phenalenyl radical.

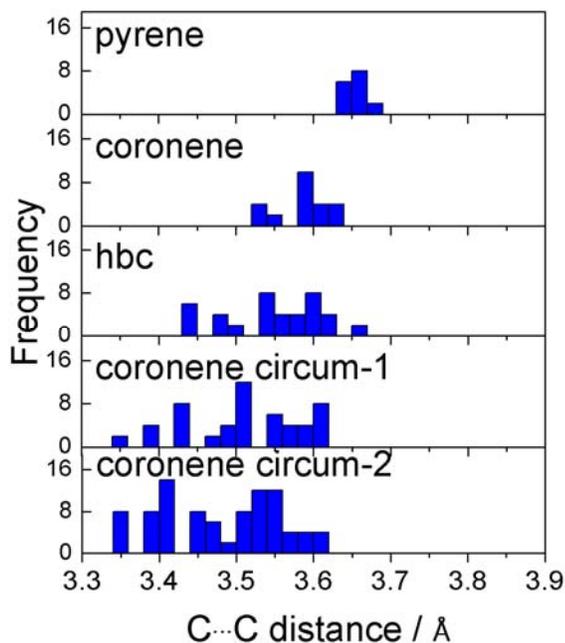

**Figure 6**. Histograms of perpendicular interring distances for different sandwich structures computed from SOS-MP2/SV(P) calculations.

3.3 Two-dimensional Sheets: Slipped-parallel Structures

Figure 7 displays the AB-type graphitic structure of the investigated slipped parallel structures. Initial tests using relative displacements of the two sheets in $C_i$ symmetry established the AB structure which was used for the purpose of computational efficiency in all following calculations by means of imposing $C_{2h}$ symmetry.

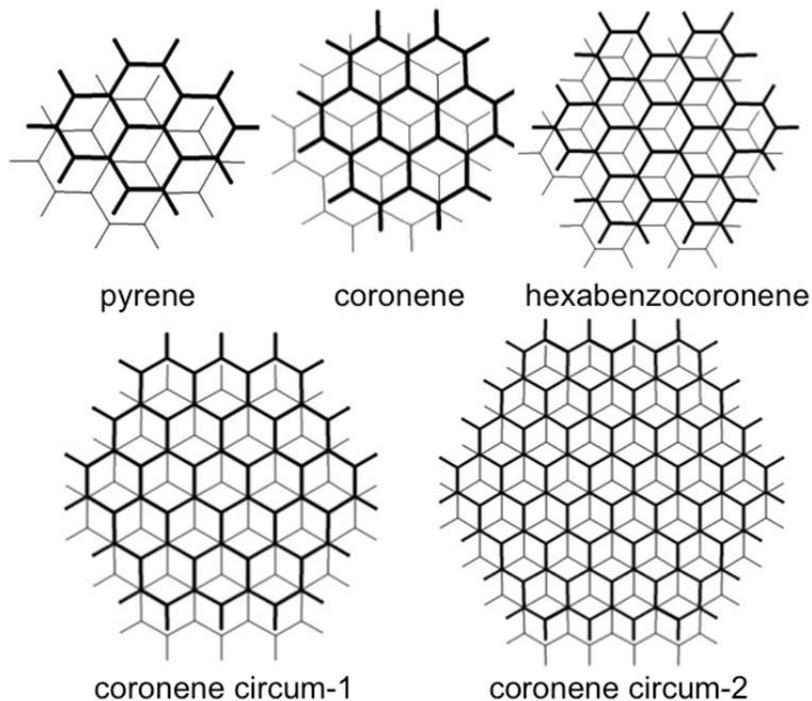

**Figure 7.** Slipped parallel structure of the two-dimensional graphene sheet dimers investigated in this work.

In Figure 8 the intermolecular interaction energies Δ*E/N* are displayed for the slipped parallel structures of pyrene to coronene circum-2 for SOS-MP2 and DFT/PBE-D3 methods. Numerical results are collected in Tables 8S to 10S. As already found for the sandwich structures, averaged SOS-MP2/SV(P) agree much better with the corresponding averaged reference def2-TZVP values (Figure 8a) which are taken as reference. Thus, they will be used in the extrapolation for $N \to \infty$ to be discussed below. The BSSE corrected DFT/PBE-D3 interaction energies are closely spaced for the two basis sets used and show a slight inversion of the order of stability for coronene circum-2. The convergence of Δ*E/N* is significantly faster in the DFT/PBE-D3 case as compared to SOS-MP2.

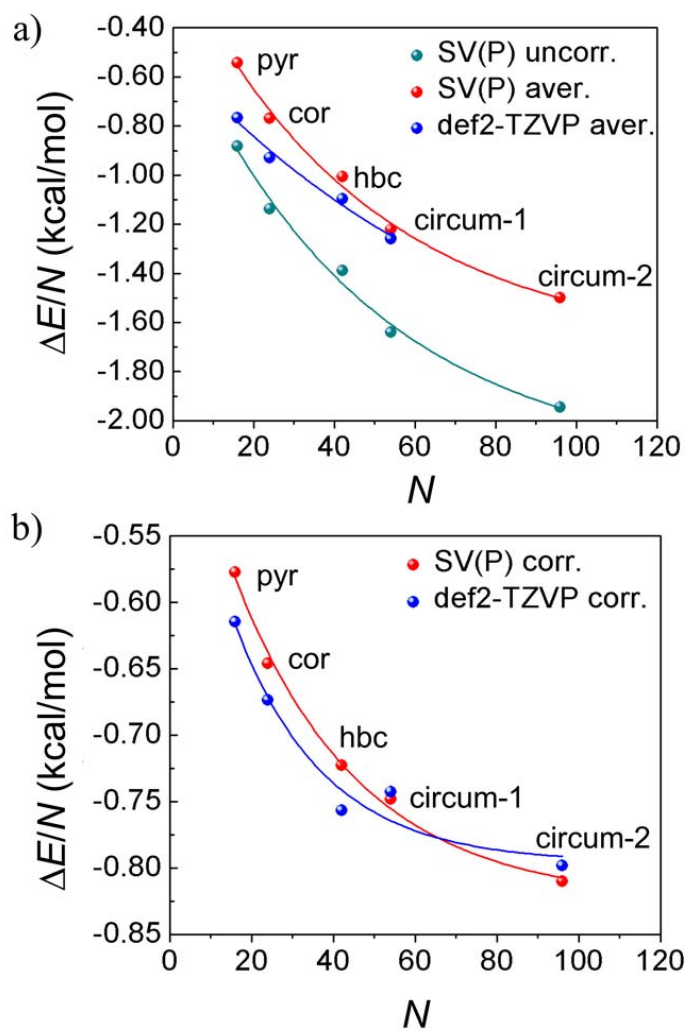

**Figure 8**. Interaction energies ($\Delta E/N$) ($N$ is the numbers of C atoms) for the slipped-parallel dimers of pyrene to coronene circum-2 in dependence of $N$ for the a) different versions of SOS-MP2 calculations and b) the BSSE corrected DFT/PBE-D3 methods.

The interaction energies $\Delta E_\infty$ extrapolated to $N \to \infty$ according to Eq. 1 are collected in Table 6. An extrapolated value of -1.71 kcal/mol (-74.3 meV) is obtained from the SOS-MP2/SV(P) calculations. This is probably slightly overbinding as compared to the complete basis set limit of SOS-MP2 as the attenuated decrease of $\Delta E/N$ for the def2-TZVP basis in comparison to the SV(P) results for the averaged interaction energies in Figure 8 indicates.

Experimental thermal desorption energy measurements of PAH's from a graphite surface[79] can be used as basis for comparison with our computed dimer association data. These measurements report an average defoliation energy of -67 meV/carbon atom which leads, after correction for interaction with hydrogen atoms, to a final value of -52 meV/carbon atom. Note that our computed interaction energies do not contain any analogous corrections for the hydrogens. In view of the above-mentioned expected basis set effects that still will reduce our computed interaction energy of -73 meV/carbon atom somewhat in absolute values, the agreement with the experimental value is quite acceptable. An interaction energy of -55 meV/atom agreeing well with the experimental value of -52 meV has been reported from quantum mechanical polarizable force field (QMPFF) simulations.[80] However, a too large interlayer separation of 3.486 Å in comparison to value of 3.35 Å[81] was computed. A somewhat lower value of -42.5 meV/atom has been reported from SAPT(DFT) calculations.[82]

**Table 6.** Extrapolated interaction energies $E_\infty$ for the two-dimensional slipped parallel structures.

| Method | $\Delta E_\infty/N$ (kcal/mol)[a] |
|---|---|
| Full optimization | |
| SOS-MP2 | |
| SV(P) aver. | -1.713 (-74.3) |
| DFT/PBE-D3 BSSE corr. | |
| SV(P) | -0.823 (-35.7) |
| def2-TZVP | -0.796 (-34.5) |
| Monomer planar, $R$ = 3.35 Å | |
| B97-D2/def2-TZVP | -1.261 (-54.7) |
| B97-D3/def2-TZVP | -1.048 (-45.5) |
| PBE-D1/def2-TZVP | -0.739 (-32.0 V) |
| PBE-D2/def2TZVP | -0.953 (-41.3) |
| PBE-D3/def2-TZVP | -0.835 (-36.2) |

[a] Values in parentheses are given in meV.

The interaction energy of -66 meV/C atom reported from DFT/B97-D calculations by Grimme et al.[25] is in good agreement with the experimental result and also with that of the present SOS-MP2 interaction energies. The situation contrasts to the strong underbinding of the DFT/PBE-D3 approach (binding energy -35 meV) found throughout this work (Figure 8 and Table 6). To explore the performance of the different dispersion interaction versions available in the Turbomole 7.01 package (-D1,[12] -D2[13] and -D3[15]), we performed calculations of the interaction energies of planar slipped parallel graphene sheet dimers constructed in the same way as in the work of Grimme et al.[25] Slipped parallel dimers of coronene up to coronene circum-3 were constructed at an intersheet distance of 3.35 Å using the monomer structures optimized at DFT/PBE-D3 level. Single point calculations were performed with the functionals B97 and PBE and the def-TZVP basis set. The results are presented graphically in Figure 9; extrapolated values are given in Table 6. These data show the strong dependence of the binding energies on the functional and version of the dispersion correction. The B97-D2 approach used by Grimme et al.[25] turns out to be by far the best combination. A similar sensitivity of DFT-based binding energies can be found also in the work of Björk et al.[24] For more investigations on the functional used on the interlayer binding in graphite see Bučko et al.[83] Since the focus of our work was laid on the exploration of *ab initio* methods we did not proceed with a more detailed search for optimal DFT procedures.

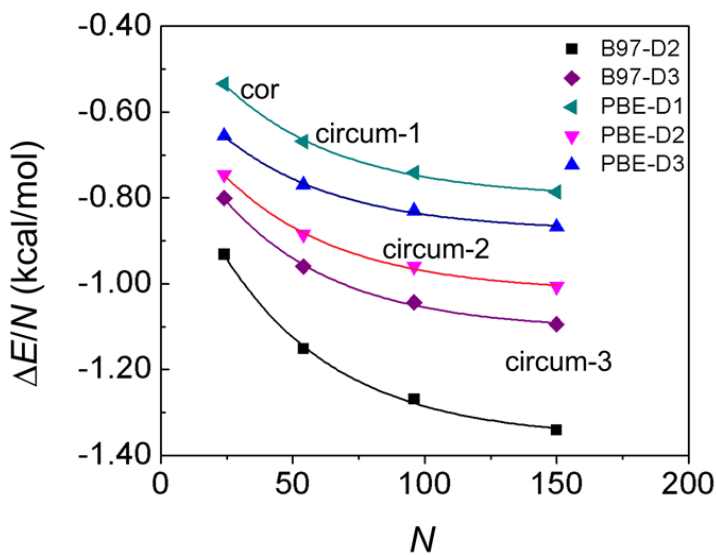

Figure 9. Interaction energies Δ*E/N* (*N* is the numbers of monomer C atoms) for the slipped-parallel dimers of coronene to coronene circum-3 using the B97 and PBE functionals, respectively, and the def2-TZVP basis.

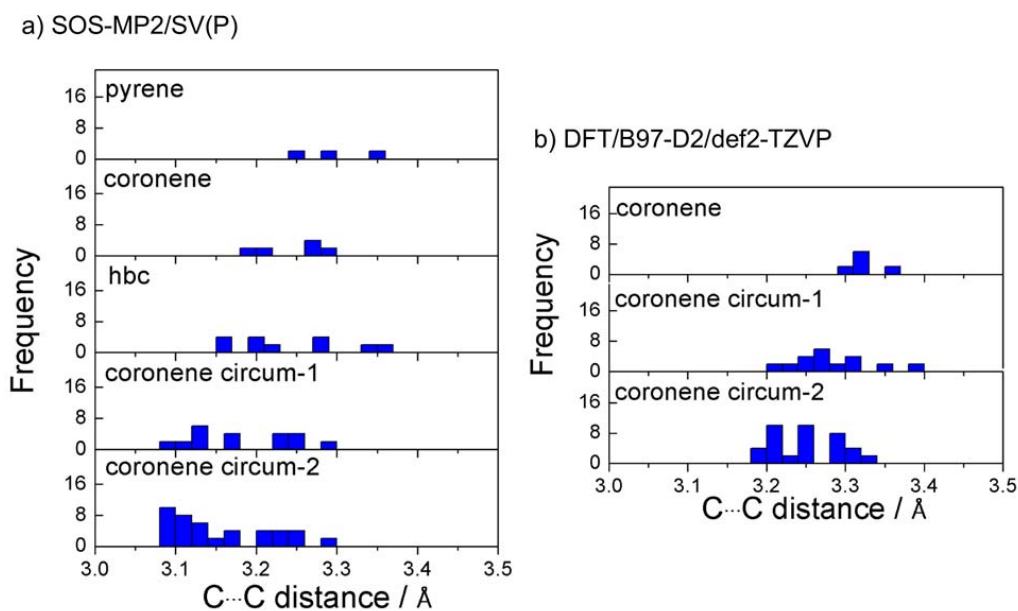

**Figure 10.** Histograms of perpendicular interring distances for different slipped parallel structures using a) the SOS-MP2/SV(P) and b) the DFT/B97-D2/def2-TZVP methods.

The out-of-plane deformations of the slipped parallel graphene sheets (Figure 10a SOS-MP2/SV(P)) and SOS-MP2/def2-TZVP – Figure 5Sa) are similarly pronounced to those obtained for the sandwich structures. The interchain C···C distances between adjacent carbon atoms vary between 3.10 Å in the inner region of the sheets and increase to ~3.30 Å at the edges (coronene circum-2). The SOS-MP2/def2-TZVP histograms shown in Figure 5Sa provide a similar picture with the onset of distances shifted by 0.1 Å to larger values. Comparable distances for the sandwich structure are in the range between 3.35 Å and 3.65 Å (Figure 6), respectively. As to be expected from the analysis of the sandwich structures, the DFT/PBE-D3 results do not show significant bending of the sheets (Figure 5Sb). From the DFT options investigated in this work only the DFT/B97-D2/def2-TZVP histograms (Figure 10b) show a slight bending.

Table 7. Stabilization energies ($\Delta E_{stab}/N$ in kcal/mol (meV in parentheses) of slipped parallel structures in comparison to the sandwich structure using the SOS-MP2 average approach.

| Structure | $\Delta E_{stab}$(SOS-MP2) SV(P) | $\Delta E_{stab}$(SOS-MP2) def2-TZVP |
|---|---|---|
| Pyrene | -0.185 (-8.03) | -0.226 (-9.80) |
| Coronene | -0.273 (-11.8) | -0.246 (-10.7) |
| HBC | -0.310 (-13.4) | -0.286 (-12.4) |
| Coronene circum-1 | -0.326 (-14.1) | -0.297 (-12.9) |
| Coronene circum-2 | -0.353 (-15.3) | - |
| Extrapolated $E_\infty$[a] | -0.342 (-14.8) | - |

[a] From extrapolation of the stabilization energies

The stabilization energy $\Delta E_{stab}/N$ per carbon atom of the slipped parallel (AB-type) vs. the sandwich (AA-type) structure is presented in Table 7. Even though experimental data for the graphitic AA structure are not available, this energy difference is considered characteristic for the description of corrugation in graphitic structures.[84] The extrapolated value of about -15 meV agrees very well with the local density approximation (LDA) and generalized gradient approximation (GGA) DFT calculations of Ref. [84] Even though interplanar binding is described quite differently in these DFT calculations (e.g. the GGA approach does not show any physically meaningful interlayer spacing) it is conjectured in the DFT work that the energy difference between AA and AB stacking structures at constant interlayer distance is still described well. It is quite gratifying that our SOS-MP2 calculations which rely on complete geometry optimizations confirm these assumptions. On the other hand, the QMPFF calculations[80] result only in 2.2 meV even though the experimental exfoliation energy is represented very well (see above).

## 4. Conclusions

Extensive computational investigations on the stacking interactions in PAHs have been performed with focus on *ab initio* methods ranging from state-of-the-art coupled cluster methods including explicitly correlated approaches to conventional Møller-Plesset theory using spin scaling. Comparison to standard DFT methods using empirical dispersion corrections is made. Quasi-linear acene dimers and two-dimensional graphene sheet models have been used; the latter range from pyrene to coronene encircled with two layers of benzene rings. The smaller dimer systems (acenes and pyrene) have been used as benchmark examples to verify the SOS-MP2 method to be applied to much larger systems where the other, higher level methods are by far out of reach. Following previous work,[59] the basis set superposition error has been taken into account within the framework of the counter-poise method in a pragmatic way using an average of the uncorrected and corrected values. Polarized triple-zeta basis sets are highly desirable, especially in view of reliable BSSE corrections, but as a compromise, polarized split-valence basis sets appear to be acceptable also. From the wide spectrum of density functionals we chose B3LYP and PBE in combination with the most recent dispersion correction D3.

Stacked sandwich structures of graphitic AA type and slipped parallel AB type structures were investigated. Under the given symmetry restraints full geometry optimizations have been performed at SOS-MP2 level. The resulting structures show a significant distortion of the PAH sheets from planarity both for AA and AB type structures. Histograms of perpendicular intersheet C···C distances show a wide spread of differences in distances within an interval of 0.2 - 0.25 Å. With 3.1 to 3.3 Å the distances are significantly smaller in the inner regions than the van der Waals distances. The finding of these biconcave structures sheds new light on the discussion of the interactions between PAH sheets showing substantial deviation from so far primarily used planar sheets. The observed bending of these sheets is, however, not so surprising in view of previously observed biconcave and biconvex dimers of substituted phenalenyl.[77, 78] On the other side, analyzing dimer structures optimized at DFT/PBE-D3 level show an insignificant out-of-plane bending of the individual sheets. Only the DFT/B97-D2 method gives a slightly bent shape of the PAH sheets in better agreement with the SOS-MP2 findings. This fact indicates, that even though it is possible to select the appropriate functional/dispersion combination to produce good intersheet binding energies, subtler, but nonetheless important, structural features are not necessarily well reproduced. These latter detailed features are, for example, important for corrugation effects and also for the discussion of structural aspects in commensurate-incommensurate transitions in graphene on hexagonal boron nitride.

Electronic supplementary information (ESI) available:

Interaction energies and intermonomer distances for stacked sandwich dimers of benzene to pentacene including BSSE corrections, results of restricted and unrestricted calculations for the stacked dimers of heptacene and decacene, interaction energies and displacement vectors for slipped parallel dimers of benzene to pentacene, interaction energies and intermonomer distances for stacked sandwich and slipped parallel dimers of pyrene to coronene circum-2 including BSSE corrections, histograms of perpendicular interring distances for different sandwich and slipped parallel structures of two-dimensional PAH sheets.

## Acknowledgments

This material is based upon work supported by the National Science Foundation under Project No. CHE-1213263 and by the Austrian Science Fund (SFB F41, ViCom). FBCM wishes to thank the Conselho Nacional de Desenvolvimento Científico e Tecnológico (CNPq) and Fundação de Amparo à Pesquisa do Estado de São Paulo (FAPESP) for financial support. We are grateful for computer time at the Vienna Scientific Cluster (VSC), project 70376, at the cluster Robinson in the TTU Department of Chemistry & Biochemistry whose purchase was funded by the National Science Foundation under the CRIFMU Grant CHE-0840493 and at the computer cluster Arran of the School of Pharmaceutical Chemistry and Technology of the Tianjin University.

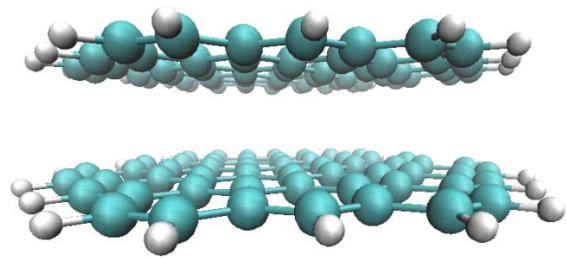

Table of Contents Figure: Slipped parallel structure of a stacked graphene flake showing a biconcave curvature.